\documentclass[fp,twocolumn]{jpsj3}
\usepackage{txfonts}
\usepackage{xcolor}

\title{Structural Phase Transition in CeMnSi under Pressure and Comparative Structural Properties of $R$MnSi ($R$ = La, Ce, Pr, Nd)}

\author{Yukihiro Kawamura$^1$\thanks{y{\_}kawamura@muroran-it.ac.jp}, Sae Nishiyama$^1$, Jun-ichi Hayashi$^1$, \\
Keiki Takeda$^1$, Chihiro Sekine$^1$, and Hiroshi Tanida$^2$}
\inst{$^1$Muroran Institute of Technology, Muroran, Hokkaido 050-8585, Japan\\
$^2$Liberal Arts and Sciences, Toyama Prefectural University, Imizu, Toyama 939-0398, Japan\\} 

\abst{
Powder X-ray diffraction experiments under pressure up to $\sim$10 GPa were performed on tetragonal CeFeSi-type $R$MnSi ($R$ = La, Ce, Pr, Nd). 
A structural phase transition was observed in CeMnSi at a critical pressure of $P_{\rm s}$ $\sim$ 5.7 GPa.
In contrast, LaMnSi, PrMnSi, and NdMnSi do not exhibit any structural transitions within the same pressure range.
The lattice parameter ratio $c/a$ of CeMnSi decreases rapidly as pressure approaches $P_{\rm s}$, whereas the $c/a$ ratios of the other $R$MnSi increase monotonically with pressure. 
CeMnSi also shows a relatively small bulk modulus: $B_0$ $\sim$ 41.4(4) GPa in the 0--2 GPa range and $B_0$ $\sim$ 32.8(2) GPa in the 4--5 GPa range, suggesting valence instability under pressure.
The structural transition in CeMnSi is attributed to the pressure-induced decrease in $c/a$ and its low bulk modulus. 
Above $P_{\rm s}$, the X-ray diffraction pattern indicates a transition to a monoclinic structure with space group No. 11, $P2_1/m$.
These findings highlight the unique pressure response of CeMnSi and provide insight into the coupling between lattice and electronic degrees of freedom in Ce-based intermetallic systems.
}

\begin{document}
\maketitle

\section{Introduction}

Crystal structure and electronic properties are closely interrelated.
In 3$d$ electron systems, certain compounds undergo a Jahn-Teller transition, which alters the crystal structure to lift the degeneracy of the $e_{\rm g}$ or $t_{\rm 2g}$ orbitals~\cite{Jahn}.
In 4$f$ electron systems, the localized nature of the 4$f$-orbitals leads to systematic lattice shrinkage, known as the Lanthanide contraction.
Even minor structural modifications can give rise to novel electronic phenomena; for example, CeSb$_2$ exhibits a unique superconducting state induced by a subtle orthorhombic distortion~\cite{Shan}.
The electronic properties of a material can influence its crystal structure, and conversely, the structure can have a profound impact on its electronic behavior.

The $RTX$ series ($R$ = rare earth, $T$ = transition element, $X$ = p-block element) comprises a group of compounds that have recently attracted attention due to the strong interplay between crystal structure and electronic properties.
$RTX$ compounds typically crystallize in the tetragonal CeFeSi-type structure (space group  $P$4$/nmm$, No. 129)~\cite{Gupta2015}.
CeCoSi exemplifies the coupling between lattice and electronic degrees of freedom, exhibiting a hidden order accompanied by a triclinic lattice distortion~\cite{Matsu2022}.
Among $R$CoSi compounds, only CeCoSi undergoes a pressure-induced structural phase transition at $P_{\rm s}$ $\sim$ 4.9 GPa~\cite{Kawa2020}.
Electrical resistivity measurements under pressure reveal a significant change in the 4$f$ electronic state associated with the structural transition, suggesting a valence instability of the Ce ion~\cite{Kawa2022}.

Recently, CeMnSi has been reported to exhibit heavy-fermion behavior originating from the Ce site, within an antiferromagnetic (AFM) state induced by the Mn sublattice~\cite{Tani2023}.
PrMnSi and NdMnSi also exhibit AFM ordering due to Mn at low temperatures, followed by a second AFM transition associated with the rare-earth elements. 
These compounds additionally undergo a structural transition around 80 K~\cite{Welter1994}.
In the $R$MnSi series, crystal structure and electronic properties appear to be strongly correlated. 
Structural refinement analysis of CeMnSi at ambient pressure suggests that it may undergo a pressure-induced structural transition, similar to CeCoSi~\cite{Tani2023}.  

It remains unclear why a pressure-induced structural phase transition occurs only in CeCoSi among $R$CoSi compounds, or whether such transitions are common in other $RTX$ systems.
The purpose of this study is to explore structural phase transitions in $R$MnSi in order to gain a comprehensive understanding of the interplay between lattice and electronic degrees of freedom in the $RTX$ system.
We have investigated the crystal structure of $R$MnSi ($R$ = La, Ce, Pr, Nd) under pressure.

Powder X-ray diffraction (XRD) results indicate that CeMnSi is the only compound among the $R$MnSi series to exhibits a structural transition at $P_{\rm s}$ $\sim$ 5.7 GPa. 
The lattice parameter ratio $c/a$ of CeMnSi shows a rapid decrease with increasing pressure above 2 GPa.
The bulk modulus $B_0$ of CeMnSi is $\sim$41.4(4) GPa in the 0--2 GPa range and $\sim$32.8(2) GPa in the 4--5 GPa range, 
which is significantly lower than those of other $R$MnSi compounds.
The observed decrease in $c/a$ and the small $B_0$ may be responsible for the pressure-induced structural transition in Ce$T$Si ($T$ = Mn, Co).
The XRD pattern suggests that the crystal structure above $P_{\rm s}$ transforms into a monoclinic structure with space group of $P2_1$/$m$ (No. 11).

\section{Experiment}
Single crystals of $R$MnSi ($R$ = La, Ce, Pr, Nd) were prepared using a self-flux method, as described in previous studies~\cite{Tani2022,Tani2023}.
The single crystals were crushed into powder for XRD measurements. Pressure was applied using a diamond anvil cell with a culet diameter of $\phi$500 $\mu$m. 
The powdered sample was pressed into a pellet with a diameter of $\sim$$\phi$150 $\mu$m and a thickness of $\sim$30 $\mu$m, and placed in a hole of a stainless-steel gasket with an inner diameter of $\phi$200 $\mu$m. 
Synchrotron X-ray with an energy of $20.00$ keV (wavelength $\lambda$ $\sim$ 0.6200 $\AA$) were used at beamline 18C of the Photon Factory at KEK, Tsukuba. The X-ray beam was collimated to a diameter of $\phi$100 $\mu$m. 
A Flat-panel detector (Rad-icon 2022, Teledyne Rad-icon Imaging Corp., USA) was used to collect the diffraction data. 
The detector resolution was 99 $\mu$m/pixel, corresponding to an angular resolution of $\Delta$2$\theta$ $\sim$ 0.02$^\circ$/pixel in this experiment. 
A 4:1 mixture of methanol and ethanol was used as the pressure-transmitting medium. 
The pressure in the sample space was calibrated using the ruby fluorescence method~\cite{Mao}.  
All XRD measurements were conducted below 10 GPa, within the pressure range where the medium remains in the liquid phase.

\section{Results}
\subsection{Structural  Transition in CeMnSi}
\begin{figure}
\includegraphics[width=\linewidth]{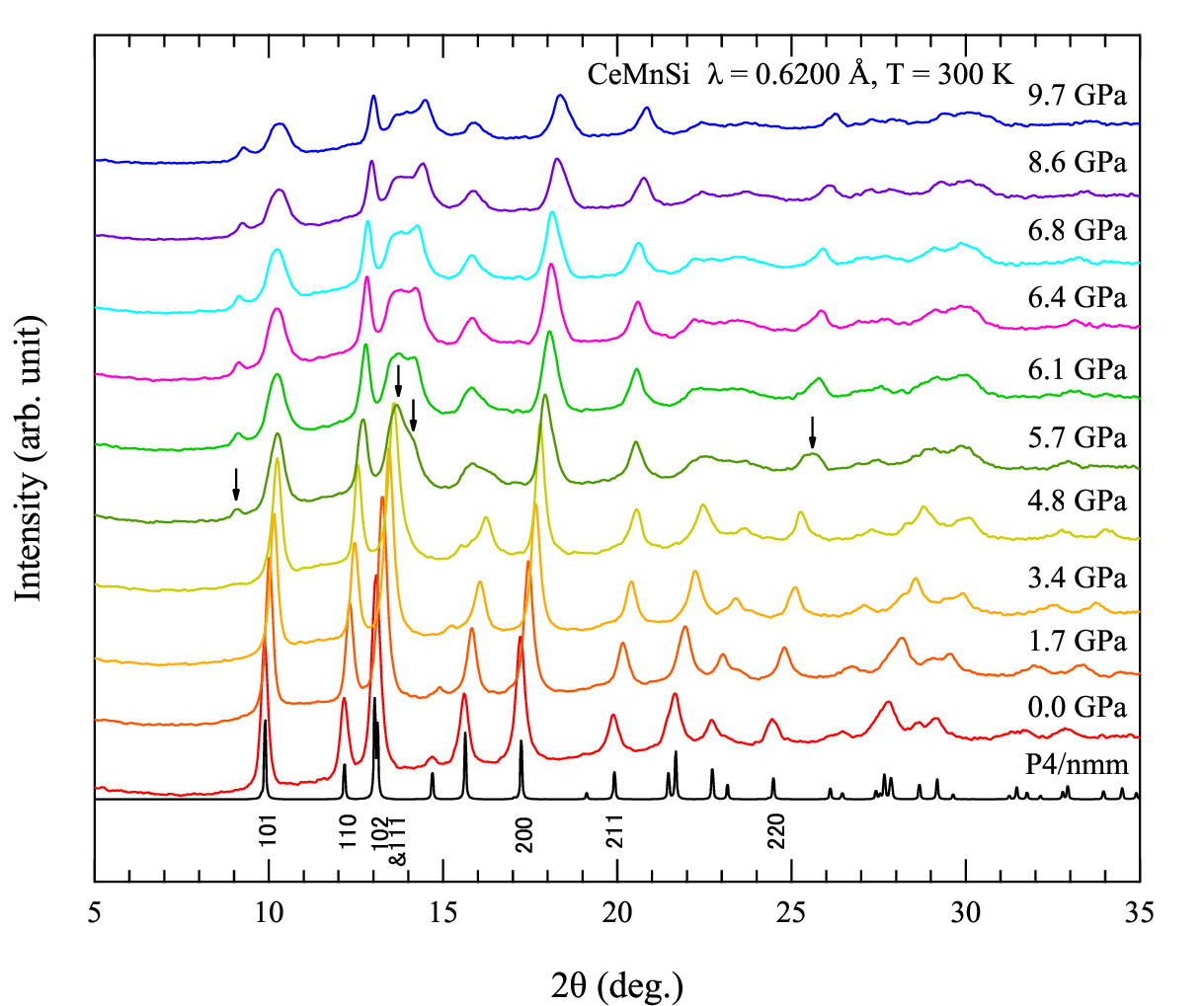}
\caption{(Color online) Powder XRD patterns of CeMnSi under pressure at 300 K, along with the calculated pattern based on the space group $P4/nmm$. Each XRD pattern is vertically offset for clarity. Arrows indicate characteristic peaks associated with the structural phase transition. }
\label{f1}
\end{figure}
Figure \ref{f1} shows powder XRD patterns of CeMnSi under pressure at 300 K, along with a simulated pattern based on the space group $P4/nmm$ shown at the bottom. 
At 0.0 GPa, all prominent peaks can be indexed to the tetragonal structure. 
The evaluated lattice parameters at 0.0 GPa are ($a$, $c$) = (4.1355(6) $\AA$, 7.272(2) $\AA$), which are consistent with those measured in laboratory conditions at ambient pressure, ($a$, $c$) = (4.127 $\AA$, 7.265 $\AA$)~\cite{Tani2023}, and with values reported for polycrystalline samples in the literature, ($a$, $c$) = (4.125(1) $\AA$, 7.285(2) $\AA$)~\cite{Welter1994}. 
The slight differences of lattice parameters ($\Delta a$, $\Delta c$) $\sim$ (0.010 $\AA$, 0.013 $\AA$) are attributed to uncertainties in the sample-to-detector distance, which was estimated from two sets of image data taken at different distances. 

The 2$\theta$ values of all prominent peaks increase monotonically with pressure up to 4.8 GPa, reflecting the continuous contraction of the unit cell.
However, at 5.7 GPa, the XRD pattern changes and can no longer be indexed using the same tetragonal crystal structure. 
A new peak appears at 9.1$^\circ$, corresponding to the (100) forbidden reflection of the $P4/nmm$ space group. 
Several other peaks also change at 5.7 GPa, as indicated by the arrows marking characteristic features in Fig. \ref{f1}. 

 \begin{figure}
\centering
\includegraphics[width=0.6\linewidth]{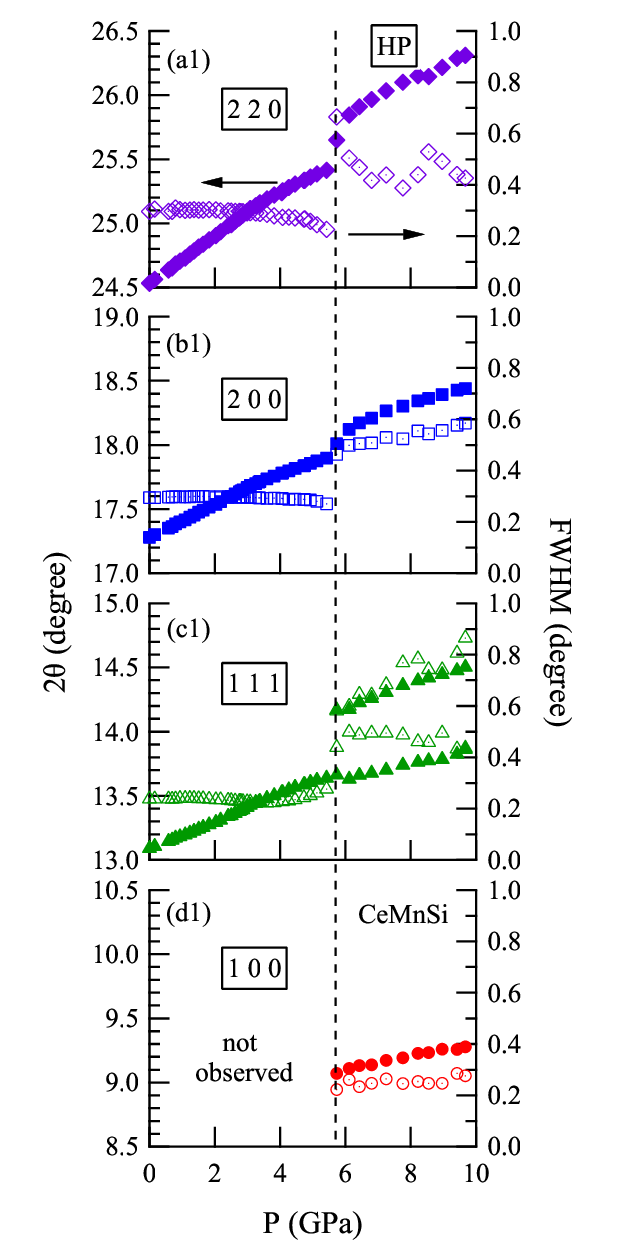}
\includegraphics[width=0.39\linewidth]{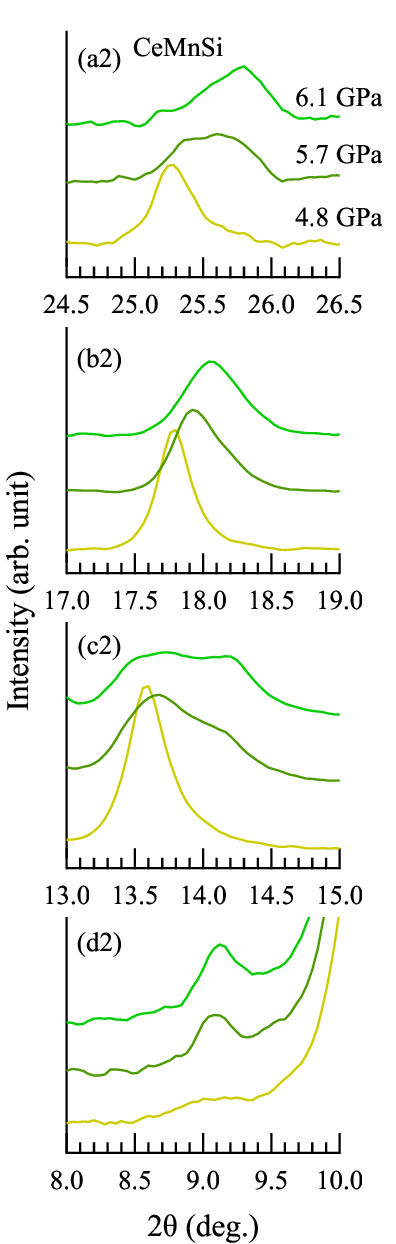}
\caption{(Color online) (a1)--(d1) Pressure dependent 2$\theta$ values (filled symbols, left axis) and FWHM (open symbols, right axis) for the diffraction index 220, 200, 111, and 100 below 5.7 GPa, and for the corresponding peaks above 5.7 GPa. 
For the peak near 2$\theta$ $\sim$ 14$^\circ$, observed above 5.7 GPa, two distinct peaks are assumed, as the profile clearly indicates the presence of multiple components.
(a2)--(d2) Enlarged views of the XRD powder patterns of CeMnSi at 300 K under pressures of 6.1 GPa (top), 5.7 GPa (middle), and 4.8 GPa (bottom).}
\label{f2}
\end{figure}
Figure \ref{f2} shows the pressure dependence of the 2$\theta$ values and the full width at half maximum (FWHM) for several diffraction peaks of CeMnSi.
Enlarged views of the XRD powder patterns at the corresponding 2$\theta$ positions are also presented.  
The peaks at approximately 2$\theta$ $\sim$13$^\circ$, $\sim$17$^\circ$, and $\sim$25$^\circ$ primarily correspond to the $P4/nmm$ diffraction indices 111, 200, and 220, respectively, and are therefore referred to as the 111, 200, and 220 peaks.
Below 5.7 GPa, the 2$\theta$ values of the 111, 200, and 220 peaks increase monotonically with increasing pressure. 
However, a significant change in their behavior is observed at 5.7 GPa.

The 2$\theta$ value of the 220 peak exhibits a discontinuous shift of approximately 0.4$^\circ$ toward higher angles at 5.7 GPa, as shown in Fig. \ref{f2}(a1).
The pressure-dependent FWHM of the 220 peak shows a maximum at 5.7 GPa.
This behavior indicates that the peak below 5.7 GPa is different from that above 5.7 GPa, and that a merged peak emerges at 5.7 GPa.
The 2$\theta$ value of the 200 peak shows a slight shift at 5.7 GPa, as illustrated in Fig. \ref{f2}(b1).
At this pressure, the FWHM of the 200 peak broadens by a factor of two.
This suggests that the 200 peak splits into multiple components above 5.7 GPa.

The 111 peak at 0 GPa is a superposition of the 102 and 111 diffraction indices of the $P4/nmm$ structure. 
This superposition appears as a single peak up to 4.8 GPa, as shown in Fig. \ref{f2}(c2).  
At 5.7 GPa, however, it evolves into a double-peak structure. 
Accordingly, the pressure dependence of the 2$\theta$ and FWHM values of the 111 peak above 5.7 GPa is plotted assuming two distinct peaks in Fig. \ref{f2}(c1).
The separation between the two peaks, denoted as $\Delta$2$\theta$, is approximately 0.5$^\circ$.
This separation cannot be reproduced by the combination of the 102 and 111 diffractions alone, suggesting the emergence of an additional peak.
Figures \ref{f2}(d1) and \ref{f2}(d2) show the appearance of the forbidden 100 reflection for the $P4/nmm$ structure, which emerges above 5.7 GPa.
These observations indicate that CeMnSi undergoes a structural phase transition at a critical pressure of $P_{\rm s}$ $\sim$ 5.7 GPa.

\begin{figure}
\centering
\includegraphics[width=\linewidth]{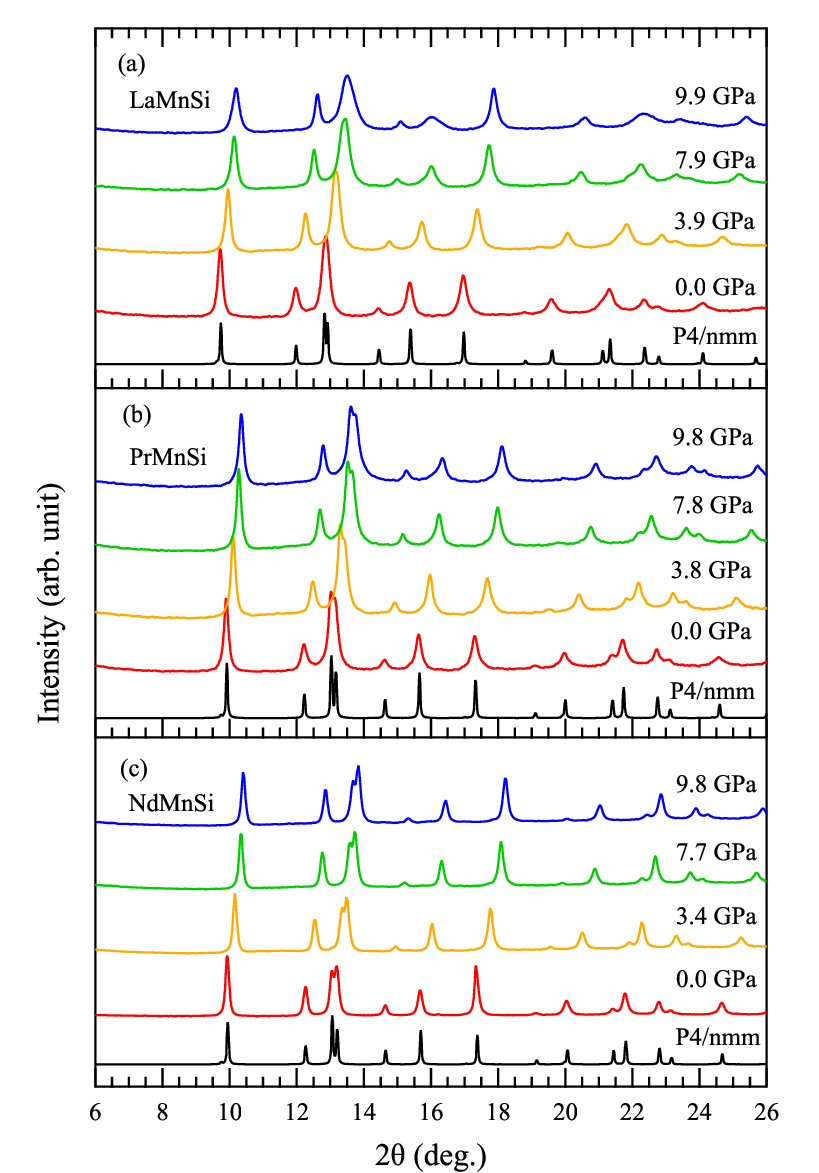}
\caption{(Color online) XRD powder patterns of (a) LaMnSi, (b) PrMnSi, and (c) NdMnSi under pressure at 300 K, along with calculated patterns based on the $P4/nmm$ space group. }
\label{f3}
\end{figure}
Figure \ref{f3} shows the XRD patterns of LaMnSi, PrMnSi, and NdMnSi under pressure.
At ambient pressure, the XRD patterns of all three compounds can be well explained by the tetragonal $P4/nmm$ space group. 
No impurity phases were detected within the resolution of this experiment.
Additionally, the 2$\theta$ values of all diffraction peaks increase monotonically with increasing pressure. 
These observations indicate that $R$MnSi ($R$ = La, Pr, Nd) does not undergo any structural phase transition below 10 GPa within the accuracy of this experiment, and that the unit cell volume decreases monotonically under pressure.
It should be noted that the peak broadening observed in the XRD pattern of LaMnSi at 9.9 GPa is attributed to the deterioration of hydrostatic pressure near the solidification pressure of the pressure-transmitting medium~\cite{Angel}.

\subsection{Pressure dependence of lattice parameters}

\begin{figure}
\centering
\includegraphics[width=0.8\linewidth]{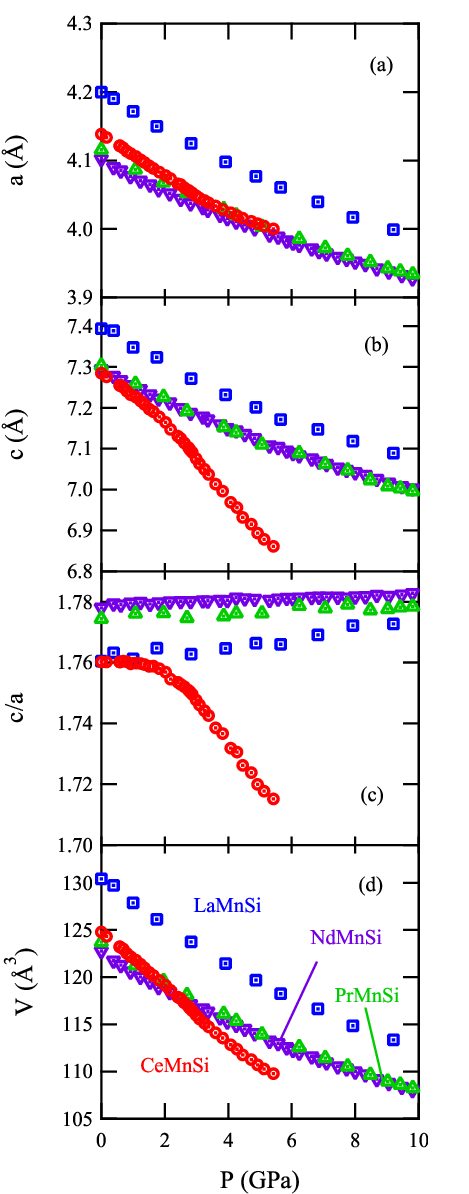}
\caption{(Color online) Pressure-dependent lattice parameters: (a) lattice constant $a$, (b) lattice constant $c$, (c) their ratio $c/a$, and (d) unit-cell volume $V$ for LaMnSi (squares), CeMnSi (circles), PrMnSi (triangles), and NdMnSi (inversed triangles) at 300 K.}
\label{f4}
\end{figure}
Figure \ref{f4} shows the pressure dependence of the lattice parameters $a$ and $c$, the $c/a$ ratio, and the unit-cell volume $V$ for $R$MnSi  ($R$ = La, Ce, Pr, Nd).
CeMnSi data are shown only below the structural transition pressure $P_{\rm s}$ $\sim$ 5.7 GPa.

At 0 GPa, LaMnSi has the largest $a$, followed by CeMnSi, PrMnSi, and NdMnSi~\cite{Tani2023}.
All compounds show a monotonic decrease in $a$ with pressure.
In CeMnSi, $a$ decreases linearly up to $\sim$3 GPa, then shows downward convexity, similar to PrMnSi.

The $c$ parameter of LaMnSi is the longest, while CeMnSi, PrMnSi, and NdMnSi have similar values at 0 GPa. 
Although $c$ decreases with pressure in all compounds, CeMnSi shows a larger pressure derivative $-dc/dP$, increasing from $\sim$0.06 \AA{}/GPa at 1 GPa to $\sim$0.09 \AA{}/GPa at 4 GPa.
As a result, $c$ for CeMnSi deviates significantly from the others, reaching $\sim$6.85 \AA{} just below $P_{\rm s}$.

The $c/a$ ratio of CeMnSi shows a more pronounced pressure dependence than $c$.
While the $c/a$ ratios of LaMnSi, PrMnSi, and NdMnSi slightly increase with pressure, that of CeMnSi remains nearly constant up to 1 GPa, then gradually decreases above 1 GPa, with a more pronounced decline beyond 2 GPa.
Overall, the $c/a$ ratio of CeMnSi decreases by approximately 3$\%$ from 1 GPa to 5 GPa. 

The unit-cell volume $V$ decreases with pressure in all compounds. 
LaMnSi, PrMnSi, and NdMnSi show a gradual suppression in the rate of decrease, while CeMnSi decreases almost linearly, with a weak kink at $\sim$3 GPa, followed by downward convexity. 
Although CeMnSi has a larger $V$ than PrMnSi and NdMnSi at 0 GPa, it becomes smaller above 3 GPa.

Inflection points in the pressure dependence of $a$, $c$, $c/a$, and $V$ around 3 GPa may suggest structural deformation.
However, the XRD pattern of CeMnSi below $P_{\rm s}$ $\sim$ 5.7 GPa remains consistent with $P4/nmm$ space group, as shown in Fig. \ref{f1}.

\section{Discussion}

Firstly, we compare the variation in the $c/a$ ratio of $RT$Si ($R$ = La, Ce, Pr, Nd, $T$ = Mn, Co) under pressure with that induced by chemical pressure resulting from lanthanide contraction. We then discuss the distinctive pressure dependence of the $c/a$ ratio in Ce$T$Si.
Secondly, we compare the bulk modulus $B_0$ of $R$MnSi compounds and highlight the notably low $B_0$ observed in CeMnSi.
Finally, based on the features of the powder XRD pattern, we propose a possible candidate for the crystal structure of CeMnSi above the transition pressure $P_{\rm s}$. 

\subsection{Pressure and chemical pressure dependence of $c/a$}

\begin{figure}[b]
\centering
\includegraphics[width=0.8\linewidth]{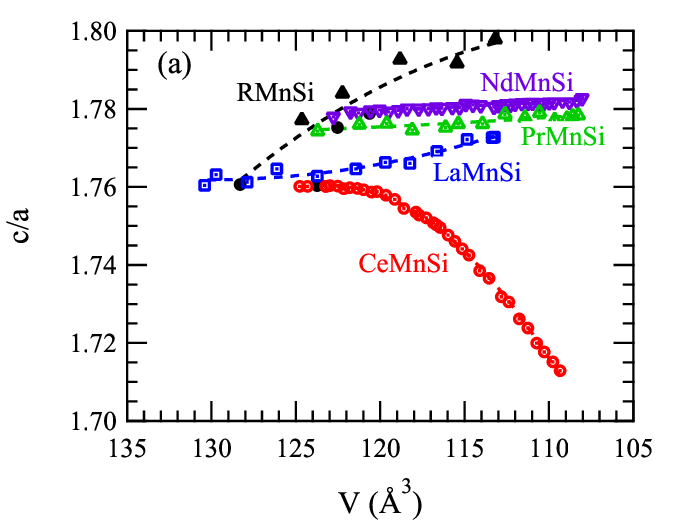}\\
\includegraphics[width=0.8\linewidth]{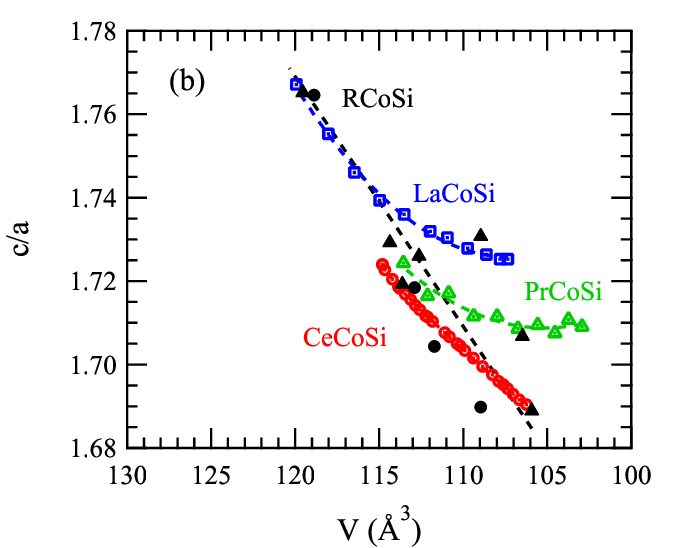}
\caption{(Color online) (a) Volume ($V$) dependence of the $c/a$ ratio in $R$MnSi ($R$ = La, Ce, Pr, Nd) under pressure (open symbols). For comparison, the $V$ dependence of $c/a$ at ambient pressure for $R$MnSi ($R$ = La, Ce, Pr, Nd, Sm, Gd, Y) reported by H. Kido et al. (filled triangles)~\cite{Kido} and by H. Tanida et al., (filled circles)~\cite{Tani2023} is also shown.  (b) Volume ($V$) dependence of the $c/a$ ratio in $R$CoSi ($R$ = La, Ce, Pr) under pressure~\cite{Kawa2022} (open symbols), along with ambient pressure data for $R$CoSi ($R$ = La, Ce, Pr, Nd, Sm, Gd, Tb) reported by O.I. Bodak et al. (filled circles)~\cite{Bodak1970} and R. Welter et al., (filled triangles)~\cite{Welter1994J}. Dashed lines are guides for the eye.}
\label{f5}
\end{figure}
The $c/a$ ratio of CeMnSi exhibits little pressure dependence up to 1 GPa, but decreases rapidly above 2 GPa.
In contrast, the $c/a$ ratios of the isostructural compounds $R$MnSi ($R$ = La, Pr, Nd) increase monotonically with increasing pressure.
To understand this behavior, we compare the pressure dependence of $c/a$ in $RT$Si ($R$ = rare earth, $T$ = Mn, Co) with its chemical pressure dependence, and discuss the origin of the marked decrease in $c/a$ at high pressures up to $P_{\rm s}$ in CeMnSi.

Figure \ref{f5}(a) shows the $c/a$ ratios of $R$MnSi ($R$ = La, Ce, Pr, Nd) plotted as a function of unit-cell volume $V$, with pressure as an implicit parameter.
For $R$ = La, Pr, and Nd, $c/a$ increases slightly as $V$ decreases. LaMnSi exhibits the most pronounced increase, and its $c/a$ approaches that of PrMnSi and NdMnSi at high pressures. This behavior may be attributed to the absence of 4$f$ electrons and/or differences in the ionic radii of the trivalent rare-earth ions. In any case, volume contraction alone has a limited effect on the $c/a$ ratio in $R$MnSi.

For comparison, the $c$/$a$ ratios of $R$MnSi ($R$ = La to Gd and Y) at ambient pressure are also shown~\cite{Kido, Tani2023}. 
A similarly weak $V$ dependence is observed as the $R$ element varies from La to Gd and Y. 
Since the ionic radius of trivalent rare-earth ions systematically decreases with increasing atomic number (known as Lanthanide contraction), substituting the $R$ element effectively applies chemical pressure. 
The similarity in $V$ dependence between chemical and external pressure suggests that the weak increase in $c/a$ is an intrinsic feature of $R$MnSi.
Even CeMnSi shows a similar weak $V$ dependence at low pressures; however, with further pressure application, $c/a$ decreases significantly until $P_{\rm s}$ = 5.7 GPa, where a pressure-induced structural transition occurs exclusively in CeMnSi.

Figure  \ref{f5}(b) presents analogous data for $R$CoSi~\cite{Kawa2020}. 
In $R$CoSi ($R$ = La, Pr), $c/a$ initially decreases with decreasing $V$ under pressure, but the rate of decrease diminishes, and $c/a$ approaches a constant value depending on the $R$ element.
The $c$/$a$ ratios of $R$CoSi at ambient pressure reported in literature~\cite{Bodak1970, Welter1994J} also decreases systematically from $R$ = La to Tb due to the chemical pressure.
These results indicate that $c/a$ in $R$CoSi decreases with decreasing $V$, whether induced by external or chemical pressure. 
In contrast, $c/a$ increases in $R$MnSi under pressure, reflecting a qualitative difference in atomic bonding between $R$MnSi and $R$CoSi~\cite{Tani2023}.
The saturation of $c/a$ may reflect a limit in lattice contraction specific to each $R$ element. Notably, in CeCoSi, $c/a$ continues to decrease until $P_{\rm s}$ = 4.9 GPa, surpassing the saturation observed in La and Pr, and a structural transition occurs exclusively in CeCoSi~\cite{Kawa2020, Kawa2022}.

As shown in Fig. 5, $c/a$ continues to decrease in both CeMnSi and CeCoSi until $P_{\rm s}$, suggesting that this behavior is not solely due to lattice effects. 
Another key factor likely governs the lattice contraction at high pressures in Ce-based compounds.
The decrease in $c/a$ implies that the lattice parameter $c$ is more compressible under pressure. The crystal structure has a quasi-two-dimensional character, with alternating $R$-$R$ and $T$-Si layers stacked along the $c$-axis.
Compression along the $c$-axis may enhance interlayer coupling, and we propose that in Ce-based compounds, regardless of $T$ = Mn or Co, the lattice contracts preferentially along the $c$-axis to maximize energy gain via $c$-$f$ hybridization.
Valence instability of the Ce ion may further promote this contraction, consistent with deviations from the Lanthanide contraction relation observed in $R$ = Ce, particularly for the lattice constant $c$~\cite{Tani2023}. 
Further pressure enhances Ce valence instability, ultimately triggering the structural phase transition at $P_{\rm s}$.

It is noteworthy that the critical pressure for the structural transition is higher in CeMnSi than in CeCoSi, although the bulk modulus is smaller in CeMnSi, as will be shown later. 
The higher $P_{\rm s}$ in CeMnSi could originate from the difference in the lattice-specific response between $R$MnSi and $R$CoSi; $c$/$a$ weakly increases in $R$MnSi whereas decreases in $R$CoSi.
The increase in $c/a$ may hinder lattice contraction, resulting in the structural transition occurring at higher pressure in CeMnSi than in CeCoSi.
A precursor behavior of the valence instability in Ce ion is suggested from the results of the electrical resistivity of CeCoSi at high pressures~\cite{EL2013, Kawa2022}. Further studies on the electrical resistivity of CeMnSi at high pressures are required, with particular interest in the pressure-induced effects on its distinctive heavy fermion state.

\subsection{$B_0$ of $R$MnSi ($R$ = La, Ce, Pr, Nd)}

\begin{figure}
\centering
\includegraphics[width=0.9\linewidth]{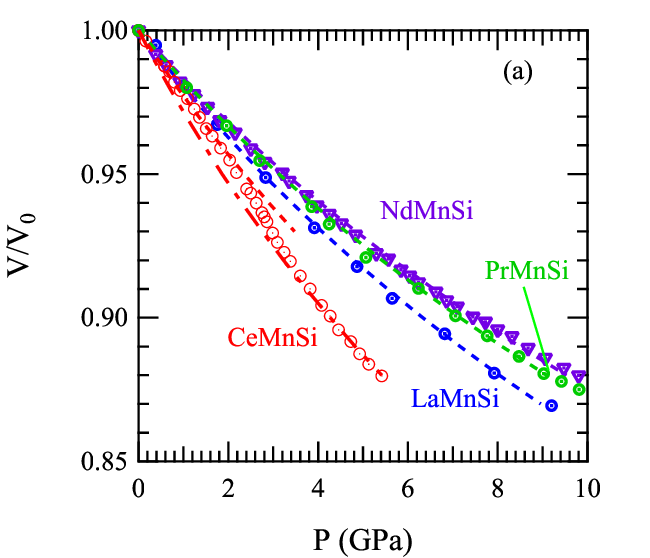}
\includegraphics[width=0.9\linewidth]{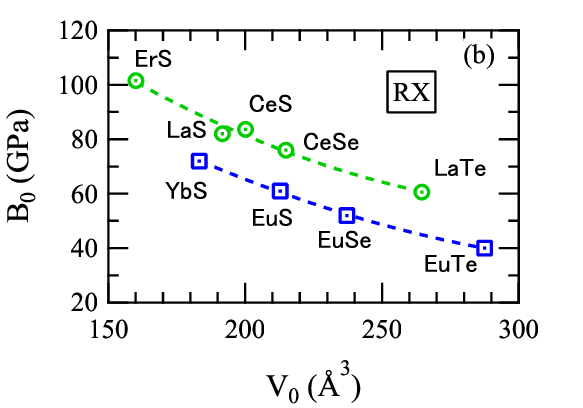}
\includegraphics[width=0.9\linewidth]{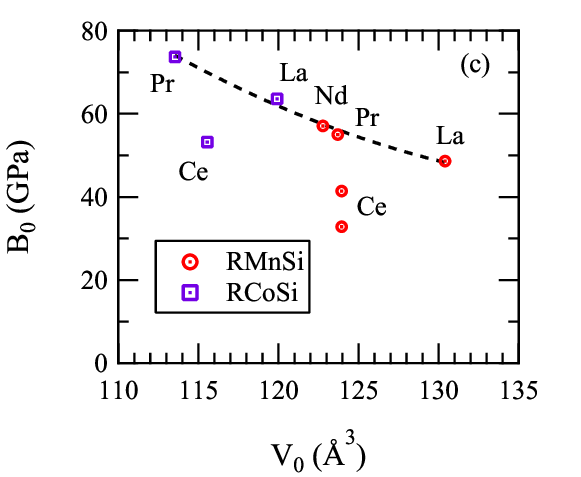}
\caption{(Color online) (a) Pressure dependence of $V/V_0$ for $R$MnSi ($R$ = La, Ce, Nd, Pr). Dashed lines for $R$MnSi ($R$ = La, Nd, Pr) represent fits to Equation (1). For CeMnSi, dashed and dash-dotted lines represent fits at 0--2 GPa and 4--5 GPa, respectively. (b) $V_0$-dependent bulk modulus $B_0$ of $RX$ ($R$ = rare earth, $X$ = chalcogen) reported by A. Jayaraman $et$ $al$.~\cite{Jaya1982}. The dashed lines are guides for the eye. (c)$V_0$-dependent bulk modulus $B_0$ of  $R$MnSi ($R$ = La, Ce, Pr, Nd) and $R$CoSi ($R$ = La, Ce, Pr)~\cite{Kawa2020}. The dashed line is a guide for the eye.}
\label{f6}
\end{figure}

Here, we discuss the bulk modulus $B_0$ of $R$MnSi ($R$ = La, Ce, Pr, Nd). 
Figure \ref{f6}(a) shows the pressure dependence of the normalized unit-cell volume $V$/$V_0$ for $R$MnSi, where $V_0$ is the volume at ambient pressure. 
Although $V/V_0$ decreases monotonically with increasing pressure for all $R$MnSi compounds, CeMnSi exhibits a notably steeper decline. 
This rapid decrease in $V/V_0$ with pressure corresponds to a smaller bulk modulus $B_0$.
The $B_0$ values were determined by least-squares fitting of the pressure-volume data to the third-order Birch-Murnaghan equation of state~\cite{Birch}.
\begin{equation}
P =\frac{3}{2}B_0 \Biggl[ \biggl(\frac{V}{V_0}\biggr)^{-\frac{7}{3}} - \biggl(\frac{V}{V_0}\biggr)^{-\frac{5}{3}} \Biggr] \Biggl\{1+\frac{3}{4}(B_0'-4)\Biggl[ \biggl(\frac{V}{V_0}\biggr)^{-\frac{2}{3}}-1\Biggr] \Biggr\},
\end{equation}
where $B_0'$ is the pressure derivative of $B_0$.
The $V/V_0$-$P$ curve of CeMnSi shows an inflection point near 3 GPa.
Therefore, $B_0$ for CeMnSi was separately calculated in the pressure range of 0--2 GPa and 4--5 GPa, as indicated by the dashed and dash-dotted lines in Fig. \ref{f6}(a), respectively.
Since $B_0'$ reflects the curvature of the $P$-$V/V_0$ curve and is highly sensitive to the fitting range, we assumed a typical value of $B_0'$ = 4. 
The resulting $B_0$ values for CeMnSi are 41.4(4) GPa in the 0--2 GPa range and 32.8(2) GPa in the 4--5 GPa range.
In contrast, the $B_0$ values for LaMnSi, PrMnSi, and NdMnSi are calculated to be 48.6(2), 55.0(2), and 57.7(2) GPa, respectively.

Empirically, $B_0$ is proportional to $V_0^{-1}$ for related compounds~\cite{Anderson}. 
For example, in rare-earth monochalcogenides $RX$ ($R$ = rare earth, $X$ = S, Se, Te) with the cubic NaCl structure, $B_0$ scales with $V_0^{-1}$,
as shown in Fig. \ref{f6}(b)~\cite{Jaya1982}.
This reflects the fact that a larger $V_0$ implies easier compression under pressure, resulting in a smaller $B_0$.
The $B_0$ values of $RX$ compounds with $R^{2+}$ and $R^{3+}$ follow different trends, which are theoretically attributed to differences in valence~\cite{Jaya1982}.
Since $R^{2+}$ has a larger ionic radius than $R^{3+}$, it is more compressible, leading to a smaller $B_0$. 
In this context, the valence of both $R^{2+}$ and $R^{3+}$ is assumed to remain constant under pressure.

Figure~\ref{f6}(c) shows the relationship between $B_0$ and $V_0$ of $R$MnSi ($R$ = La, Ce, Pr, Nd) and $R$CoSi ($R$ = La, Ce, Pr)~\cite{Kawa2020}.
Two key features are observed. 
First, $B_0$ decreases with increasing $V_0$. 
For example, LaCoSi has a larger $V_0$ and a smaller $B_0$ than PrCoSi, and similarly, LaMnSi has a larger $V_0$ and a smaller $B_0$ than PrMnSi and NdMnSi have. 
For each $R$, $R$MnSi shows a larger $V_0$ and a smaller $B_0$ than $R$CoSi.
This trend resembles the $B_0$-$V_0$ relationship observed in the $RX$ series.
Second, Ce compounds exhibit unusually small $B_0$. 
In the $RT$Si system, $B_0$ generally decreases with increasing $V_0$, except for CeCoSi and CeMnSi.  
These Ce compounds appear to undergo enhanced volume contraction under pressure, resulting in lower $B_0$.
This behavior may be attributed to pressure-induced shrinkage of the ionic radius. 
In rare-earth intermetallics, Ce is known to exhibit valence fluctuations. 
The anomalously small $B_0$ values of CeCoSi and CeMnSi suggest valence instability of the Ce-4$f$ electrons.

\subsection{Crystal structure above $P_{\rm s}$ of CeMnSi}

We observed a structural transition in CeMnSi at the critical pressure $P_{\rm s}$ $\sim$ 5.7 GPa. 
Several features in the powder XRD pattern were used to determine the crystal structure under high pressure (CSHP) above $P_{\rm s}$.
Firstly, a peak appears at 2$\theta$ $\sim$ 9$^\circ$, corresponding to the forbidden 100 reflection for the space group $P4/nmm$.
The emergence of this peak indicates that CSHP does not obey the $h00$ extinction rule, suggesting a change in symmetry.
Secondly, although some peaks broaden or shift, the overall powder XRD pattern remains similar to that of $P4/nmm$.
This implies that the crystal structure above $P_{\rm s}$ has lower symmetry than $P4/nmm$, and CSHP may belong to one of its subgroups.
Thirdly, the orthorhombic structure fails to reproduce the broad and complex peak at 2$\theta$ $\sim$ 14$^\circ$, especially at 9.7 GPa. 
This suggests that the symmetry must be further reduced to monoclinic. 
Fourthly, the full width at half maximum (FWHM) of the 101 reflection increases by 87$\%$ above $P_{\rm s}$, while that of the 110 reflection increases by only 13$\%$.
This selective broadening indicates peak splitting. The significant increase in FWHM of 101, compared to the modest change in 110, implies a change in the interaxial angle $\beta$ rather than $\gamma$.
Finally, due to the reduced symmetry, the 101 reflection splits into 101 and 011, and the 200 reflection splits into 200 and 020.
As shown in Fig. 1, the FWHM broadening of the former is 18$\%$ larger than that of the latter. 
This suggests a substantial deviation of $\beta$ from 90 $^\circ$, while the difference between $a$ and $b$ remains relatively small.

\begin{figure}
\centering
\includegraphics[width=\linewidth]{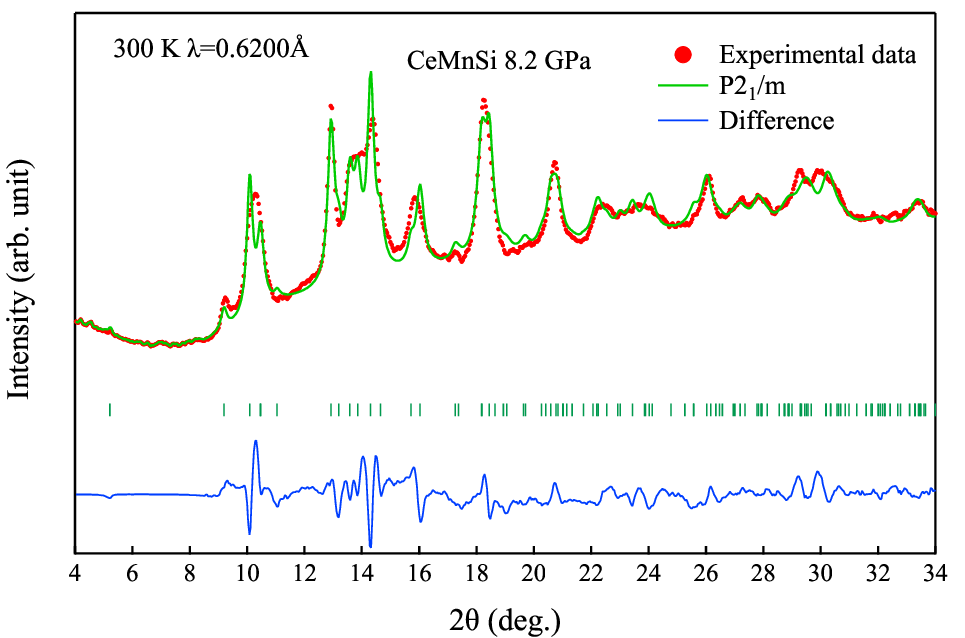}
\caption{(Color online) Powder XRD pattern of CeMnSi at 8.2 GPa ($>$ $P_{\rm s}$) and results of Rietveld refinement assuming the space group $P2_1/m$.}
\label{f7}
\end{figure}

\begin{figure}
\centering
\includegraphics[width=\linewidth]{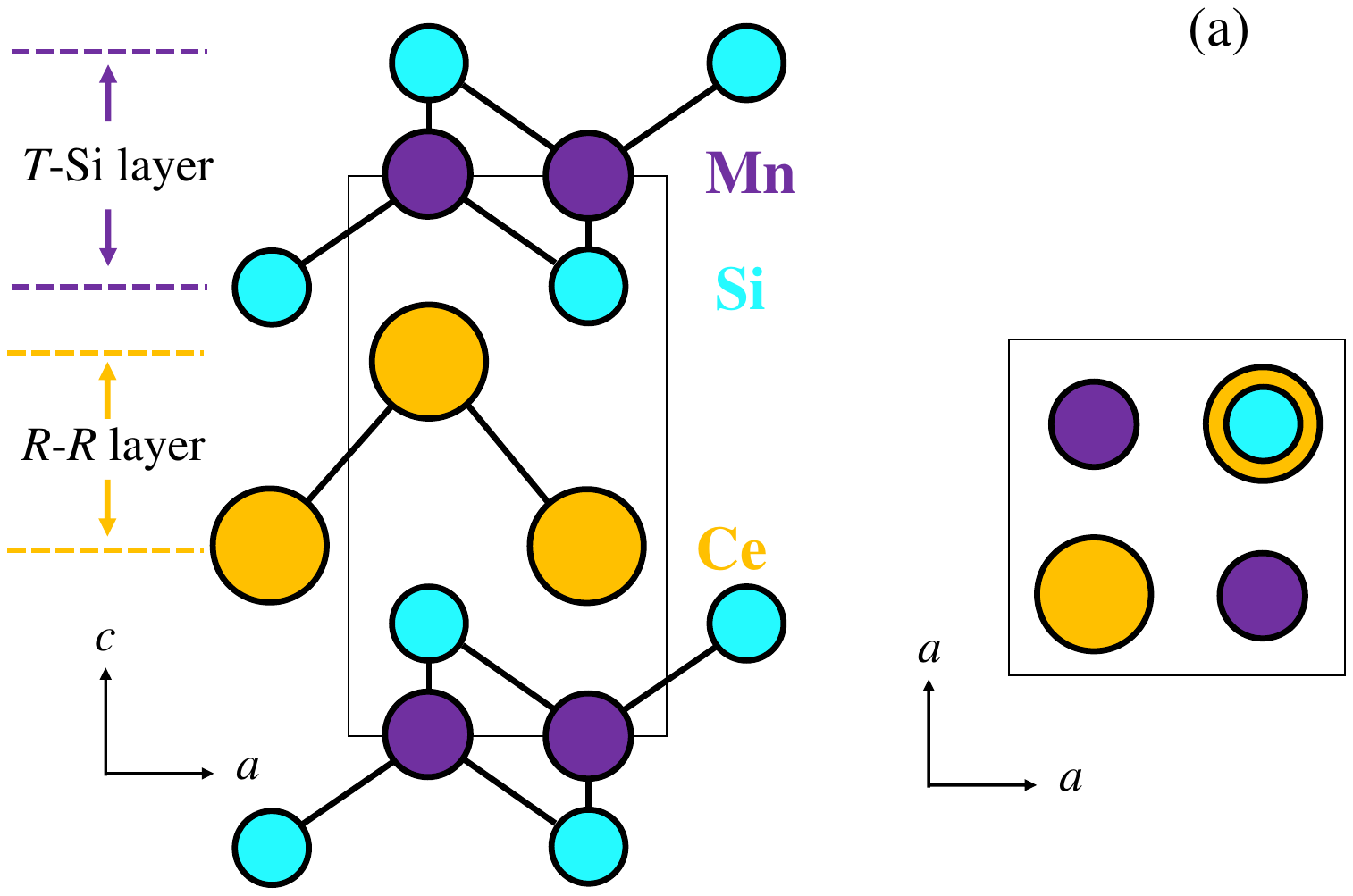}
\includegraphics[width=\linewidth]{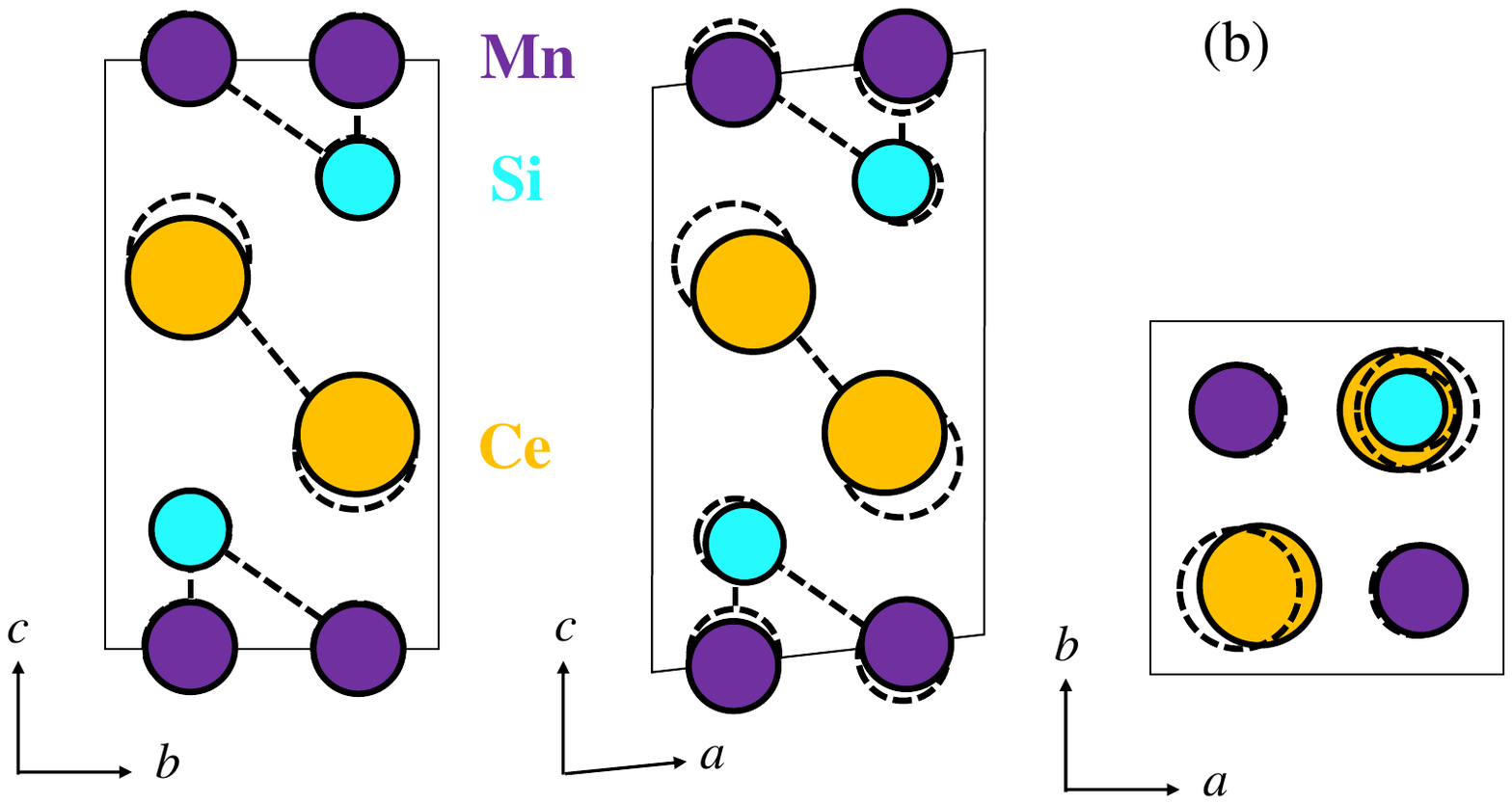}
\caption{(Color online) Crystal structure of CeMnSi: (a) low-pressure phase (space group No. 129, $P$4/$nmm$) and (b) a possible high-pressure phase candidate (space group No. 11, $P$2$_1$/$m$).}
\label{Fcrystal}
\end{figure}

Considering the five conditions discussed above, the most likely crystal structure above $P_{\rm s}$ is monoclinic with space group $P2_1/m$ (No. 11).
One of the maximal non-isomorphic subgroups (MNIS) of $P4/nmm$ is $Pmmn$ (No. 59). $P2_1/m$ is one of the MNIS of $Pmmn$.
The detailed transition from $P4/nmm$ to $P2_1/m$ is described in the supplementary material in Ref. \citen{Matsu2022}.
Figure \ref{f7} shows the results of Rietveld refinements of CeMnSi at 8.2 GPa, assuming the space group $P2_1/m$.
The refinement was performed using RIETAN-FP~\cite{RIETAN}, and the refined atomic coordinates are listed in Table I(b).
The atomic coordinates $x$ and $y$ for each atom in $P2_1/m$ differ from those in the original $P4/nmm$ structure.
Among them, the $x$ coordinate of Ce shows the largest shift, with a value of 0.0506(11).
The lattice parameters are listed in Table II.
Figure \ref{Fcrystal} shows the crystal structure of CeMnSi, drawn using VESTA~\cite{VESTA}. In the $P2_1/m$ space group, two Ce atoms move closer to each other in both the $x$ and $z$ directions compared to the $P$4/$nmm$ structure.  
The refinement parameters from the Rietveld analysis are $R_{wp}$ =  3.30 and $S$ = 2.78, indicating that there is still room for improvement.
In the calculation, the diffraction patterns around 2$\theta$ $\sim$ 10$^\circ$ and 13$^\circ$ show sharper peaks and require a greater number of peaks to achieve a good fit. 
Although long-period structures may need to be considered, it is difficult to refine the pattern using only powder XRD data.
To accurately identify the CSHP, single-crystal XRD analysis is desirable, and such studies are currently in progress.
As a supplementary note, the XRD pattern of CeCoSi exhibits features similar to those of CeMnSi, such as the appearance of the forbidden 100 reflection and the splitting of the 111 peak. Although the high-pressure phases (CSHP) of CeCoSi and CeMnSi are likely to share the same space group, the Rietveld refinement for CeCoSi, like that for CeMnSi, still leaves room for improvement.

\begin{table}[h]
\caption{Atomic coordinates of CeMnSi: (a) low-pressure phase (space group No. 129, $P$4/$nmm$) at ambient pressure~\cite{Tani2023}, and (b) a candidate high-pressure phase (space group No. 11, $P$2$_1$/$m$) at 8.2 GPa.}
\label{t1}
\begin{tabular}{cccccc}
(a) $P$4/$nmm$ & & & & & \\
\hline
atom & site & Symmetry & $x$ & $y$ & $z$\\
\hline
Ce & 2$c$ & 4$mm$ & 1/4 & 1/4 & 0.664\\
Mn & 2$a$ & $\bar{4}m$2 & 3/4 & 1/4 & 0\\
Si & 2$c$ & 4$mm$ & 1/4 & 1/4 & 0.200\\
\hline
\\
(b) $P$2$_1$/$m$ & & & & & \\
\hline
atom & site & Symmetry & $x$ & $y$ & $z$\\
\hline
Ce & 2$e$ & $m$ & 0.3006(11) & 1/4 & 0.6327(16)\\
Mn & 2$e$ & $m$ & 0.760(9) & 1/4 & 0.003(8)\\
Si & 2$e$ & $m$ & 0.279(17) & 1/4 & 0.204(11)\\
\hline
\end{tabular}
\end{table}

\begin{table}[h]
\caption{Lattice parameters of CeMnSi in the low-pressure phase at 3.2 GPa (space group No. 129, $P$4/$nmm$) and in the candidate high-pressure phase at 8.2 GPa (space group No. 11, $P$2$_1$/$m$).}
\label{t2}
\begin{tabular}{ccccc}
\hline
Pressure & a $(\AA)$ & b $(\AA)$ & c $(\AA)$ & $\beta$ $(^\circ)$ \\
\hline
3.2 GPa & 4.0415(5) & 4.0415(5) & 7.048(2) & 90\\
8.2 GPa & 3.8849(13) & 3.9170(16) & 6.814(6) & 83.78(4)\\
\hline
\end{tabular}
\end{table}

\section{Summary}

Powdered XRD on $R$MnSi ($R$ = La, Ce, Pr, Nd) under pressure have been performed.
Among $R$MnSi ($R$ = La, Ce, Pr, Nd) compounds, only CeMnSi exhibits a structural phase transition at $P_{\rm s}$ $\sim$ 5.7 GPa. 
The lattice parameter ratio $c/a$ of CeMnSi decreases rapidly above 2 GPa, 
whereas that of LaMnSi, PrMnSi, and NdMnSi increases monotonically with pressure.
This decrease in $c/a$ is attributed to the anisotropic $c$-$f$ hybridization, which is considered one of the driving forces behind the structural phase transition.
Additionally, CeMnSi shows an anomalously low bulk modulus $B_0$ among the $RT$Si ($T$ = Mn, Co) compounds, suggesting gradual valence instability of the Ce ion, similar to that observed in CeCoSi. 
The low $B_0$ is also regarded as a contributing factor to the structural transition.
X-ray diffraction pattern above $P_{\rm s}$ indicates that the high-pressure phase most likely adopts a monoclinic structure with space group No.11, $P$2$_1$/$m$.


\section*{Acknowledgment}

Synchrotron X-ray diffraction was performed at KEK BL-18C, with the approval of the Photon Factory Program Advisory Committee (Proposal No. 2023G532).
A portion of this study was supported by JSPS KAKENHI
Grant Nos. JP22K19076, JP25K01485, JP25K07209.

\end{document}